\documentstyle{article}

\newcommand{\mathbf}{\bf}

\begin{document}

\begin{center}
{\huge\bf Uncertainty Relations for Two Dimensional Quantized  
Electromagnetic Potential}
\end{center}

\vspace{1cm}
\begin{center}
{\large\bf
F.GHABOUSSI}\\
\end{center}

\begin{center}
\begin{minipage}{8cm}
Department of Physics, University of Konstanz\\
P.O. Box 5560, D 78434 Konstanz, Germany\\
E-mail: ghabousi@kaluza.physik.uni-konstanz.de
\end{minipage}
\end{center}

\vspace{1cm}

\begin{center}
{\large{\bf Abstract}}
\end{center}

\begin{center}
\begin{minipage}{12cm}
The canonical quantization of flux is performed. It is shown that  
according to the canonical flux quantization there must be a new  
uncertainty relation: $e \Delta A_m . \Delta x_m \geq \hbar$ where  
$A_m$ and $\Delta x_m \geq l_B$ are the electromagnetic gauge  
potential, the position uncertainty and the magnetic length,  
respectively. Other arguments in favour of this uncertainty relation  
are also discussed.

\end{minipage}
\end{center}

\newpage
The flux quantization is described according to the relation:  
$\oint e A_m dx^m = \int \int e F_{mn} dx^m \wedge dx^n = \Phi  = N  
h , N \in {\mathbf Z}$ where $A_m$ and $F_{mn}$ are the  
electromagnetic potential and the magnetic field strength and $m, n  
= 1, 2$. We prove that in this case there must be a new uncertainty  
relation $e \Delta A_m . \Delta x_m \geq \hbar$ which is equivalent  
to the canonical flux quantization according to the quantum  
commutator postulate $e [ \hat{A}_m  ,  \hat{x}_m ] = -i \hbar$.  
Hereby $x_m$ can be considered either as the coordinates of the  
centre of cyclotron motion or as the relative coordinates around the  
centre \cite{qml}.

To begin we show that such a commutator can be considered as a  
result of electronic behaviour in magnetic fields:

From the usual requirement in flux quantization that the electronic  
current density $j_m = ne \hat{V}_m = \Psi^* (\hat{p}_m - e  
\hat{A}_m) \Psi$ must vanish in the region where the contour  
integral $\oint A_m dx^m$ takes place \cite{current}, one concludes  
that {\it in this region} $[ \hat{V}_m \ , \ \hat{x}_m ] = 0$. This  
implies that in this region $[\hat{p}_m \ , \ \hat{x}_m ] = e [  
\hat{A}_m \ , \ \hat{x}_m ]$ or that $ e [ \hat{A}_m s , \ \hat{x}_m  
] = -i \hbar$.

\medskip
Moreover, it is also known that for the cyclotron motion of  
electrons, the coordinate operators of relative coordinates are  
non-commuting. Thus, one has $[\hat{x}_m \ , \ \hat{x}_n] = -i l_B  
^2 \epsilon_{mn}$ for the relative cyclotron coordinates, where  
$l_B$ is the magnetic length \cite{qml}. This is an interesting  
example of the non-commutative geometry of configuration space in  
quantum theory. Now, the mentioned commutator $[ \hat{A}_m \ , \  
\hat{x}_m]$ is proportional to this commutator in the usual Landau  
gauge $A_m = B x^n \epsilon_{mn} \ , \epsilon_{mn} = - \epsilon_{nm}  
= 1$ which was introduced to study the behaviour of electrons in  
magnetic fields \cite{qml}. Therefore, in view of this  
proportionality one has indeed $[ \hat{A}_m \ , \ \hat{x}_m] = B  
\epsilon_{mn} [\hat{x}_n \ , \ \hat{x}_m] = -i l_B ^2 \cdot B = -i  
\displaystyle{\frac{\hbar}{e}}$ for $C = 1$.

\medskip
It should be mentioned also that the usual argument, that the  
electromagnetic potential $A_m$ is a function of $x^m$ and therefore  
the operators $\hat{A}_m$ and $\hat{x}_m$ must commute with each  
other, does not apply to the case of flux quantization:

Here $A_m$ is not a function of $x^m$, but it is given either by  
$A_m = B \cdot x^n \epsilon_{mn}$ {\it within} the flux surface  
where $\epsilon_{mn} F_{mn} (A_m)_{(surface)} = B$ is constant, or  
it is given by an electromagnetic pure gauge potential $\tilde{A}_m  
:= \partial_m \phi$ in the contour region \cite{current}.
In these both related cases which are relevant for the flux  
quantization the electromagnetic potential $A_m$ is not a function  
of $x_m$. Therefore, there is no a periori reason for the  
commutativity of the operators $\hat{A}_m$ and $\hat{x}_m$.

After these consistency arguments for the non-triviality of  
mentioned commutator and uncertainty relations we give a more  
rigorous prove for their existence according to the canonical  
quantization structure of the flux action functional.

\medskip
We will show that, indeed for the true phase space variables of the  
two dimensional electromagnetic system of flux quantization, the  
commutator of related operators is non-trivial and so there exist  
equivalent uncertainty relations. The key point is the choise of  
correct phase space, i. e. the choise of true canonical conjugate  
variables for the electromagnetic system under consideration, which  
has to be quantized in order to describe the flux quantization.

The point of departure is the flux quantization relation for  
electromagnetic system:

\begin{equation}
\int \int e F_{mn} dx^m \wedge dx^n = \Phi = N h
\end{equation}
\label{one}

This quantization should be, in principle, describable by the  
canonical quantization of the classical action functional $S_{(Cl)}  
^{(flux)}$ which is given naturally by the flux quantization  
relation (1):

\begin{equation}
S_{(Cl)} ^{(flux)} = \int \int e F_{mn} dx^m \wedge dx^n = \int  
\int e dA_n \wedge dx^n \ ,
\end{equation}
\label{two}

where we used $ dA_n := \partial_m A_n dx^m$.

To quantize the phase space of a classical system which is  
represented by an action functional $S_{(Cl)}$, one should determine  
first the canonical conjugate variables of phase space and then one  
should postulate the quantum commutator for operators which are  
related to these variables. Now to determine the phase space space  
variables of the system which is represented by the action  
functional $S_{(Cl)}$ one should compare it with the canonical  
action functional:

\begin{equation}
S_{(Cl)} ^{(canon)} = \int \int dp_m \wedge dx^m \ ,
\end{equation}
\label{four}

of the same dimension \cite{comp}.

The comparison between $S_{(Cl)} ^{(flux)}$ in (2) and  
$S_{(canon)}$ in (3) shows that the phase space of our system which  
is represented by $S_{(Cl)} ^{(flux)}$ has the set of canonical  
conjugate variables: ${\{ e A_m, x^m}\}$.

\bigskip
Then, the globally Hamiltonian vector fields of our system which  
has the symplectic 2-form

$\omega = e dA_n \wedge dx^n = e F_{mn} dx^m \wedge dx^n$ are given  
by the following differential operators \cite {wood}, \cite{erk}:

\begin{equation}
X_{A_m} = \displaystyle{\frac{\partial}{\partial x^m}} \;\;\; ,  
\;\;\; X_{x^m} = - \displaystyle{\frac{\partial}{\partial A_m}}
\end{equation}
\label{five}

Moreover, the quantum differential operators on the quantized phase  
space of this system should be proportional to these vector fields  
by a complex factor, i. e. usually by $(-i \hbar )$, and so they  
should be given by $\hat{A} = -i \hbar  
\displaystyle{\frac{\partial}{\partial x^m}}$ and  $\hat{x} = i  
\hbar \displaystyle{\frac{\partial}{\partial A_m}}$.

On the other hand, the actual quantized phase space of a quantum  
system should be polarized in the sense that the wave function of  
the system should be a function of only half of the variables of the  
original phase space \cite{wood}. This means that in general $\Psi$  
is either in the $\Psi ( p_i , t)$- or in the $\Psi (x^i , t)$  
representation. Then, the half of quantum operators which are  
related to the variables in $\Psi$ act on $\Psi$ just by the  
multiplication with these variables and the second half of quantum  
operators act on it by the action of quantum differential operators  
discussed above. In other words, as it is well known, for example in  
the $\Psi ( p_i , t)$ representation the acting operators are given  
by $\hat{x}^i = - i\hbar X_{x^i}^{(canon)} = i \hbar  
\displaystyle{\frac{\partial}{\partial p_i}}$ and $\hat{p}_i = p_i$,  
which have the correct commutators: $[\hat{p}_i \ , \ \hat{x}^j ] =  
-i \hbar \delta_i ^j$. The same is true also in the $\Psi (x^i ,  
t)$ representation for the $\hat{x} = x$ and $\hat{p}_i = -i \hbar  
\displaystyle{\frac{\partial}{\partial x^i}}$ operators.

In our case where in view of the neccessary polarization the wave  
function of ${\{A_m , x^m}\}$ system is either in $\Psi ( A_m , t)$  
or in $\Psi ( x^m , t)$ representation, the quantum operators are  
given either by the set ${\{ \hat{A}_m = A_m \ , \ \hat{x}_m = -i  
\hbar X_{x^i} =
i \hbar \displaystyle{\frac{\partial}{\partial A_m}}}\}$ or by the  
set ${\{ \hat{A}_m = -i \hbar X_{A_m} = -i  
\hbar\displaystyle{\frac{\partial}{\partial x^m}} \ , \ \hat{x}^m =  
x^m }\}$, respectively.

In both representations the commutator between the quatum operators  
is given by $(-i \hbar)$:

\begin{equation}
e [ \hat{A}_m \ , \ \hat{x}_n ] \Psi = -i \hbar \delta_{mn} \Psi
\end{equation}
\label{six}

Equivalently, we have according to quantum mechanics a true  
uncertainty relation for $A_m$ and $x_m$, i. e.: $e \Delta A_m \cdot  
\Delta x_m \geq \hbar$.

In other words, to describe the flux quantization according to the  
canonical quantization sheme, one has to consider the commutators  
(5) and also the equivalent uncertainty relations $e \Delta A_m  
\cdot \Delta x_m \geq \hbar$.

In the Landau gauge there should be also an equivalent uncertainty  
relation which is given by:

$e B \Delta x_m . \Delta x_n \geq \hbar |\epsilon_{mn}|$ , i. e.  
for $m \neq n$. This uncertainty relation is related to the $e B [  
\hat{x}_m \ , \hat{x}_n ] = -i \hbar \epsilon_{mn}$ commutator. The  
same uncertainty relation can be obtained also from the original  
uncertainty relations, if we use the Landau gauge $\Delta A_m = B .  
\Delta x^n \epsilon_{mn}$.

\bigskip

Moreover, the electromagnetic gauge potential have according to the  
uncertainty relations

$e \Delta A_m \cdot \Delta x_m \geq \hbar$ a maximal uncertainty of  
$(\Delta A_m)_{max} = \displaystyle{\frac{\hbar}{e l_B}}$ for the  
case $ \Delta x_m = l_B$. Hence, using the Landau gauge one obtains  
the definition of magnetic length $ l^2 _B =  
\displaystyle{\frac{\hbar}{e B}}$ which proves the consistency of  
this approach.

\bigskip

Footnotes and references


\begin{thebibliography}{100}
\bibitem{current}

In the usual derivation of flux quantization the integration path  
is chosen in a region where the electronic current density is zero.  
The vanishing of current density in a region, i. e. here in the  
contour region where the integral $\oint A_m dx^m$ is performed, is  
equivalent to the definition of the electromagnetic potential as a  
pure gauge potential in this region, i. e. $A_{m(contour)} =  
\tilde{A}_m \ ,$

$F_{mn} (\tilde{A}_m )_{(contour)} = 0$.
\bibitem{comp}
This procedure of canonical comparison to identify the phase space  
variables of a new system for its quantization is the very general  
and usual one which is used for any canonical quantization, e. g.  
for the quantization of the $(3+1)$ decomposed Yang-Mills action  
functional, Maxwell's electrodynamics, the $(2+1)$ decomposed  
Chern-Simons action functional, etc.
\bibitem{erk}
According to the geometric quantization \cite{wood} the classical  
Hamiltonian vector fields related to the canonical conjugate  
variables of our ${\{ A_m , x^m }\}$ phase space are given in  
general by:

$X_{A_m} = \displaystyle{{\frac{\partial A_m}{\partial  
A_n}}{\frac{\partial}{\partial x^n}} - {\frac{\partial A_m}{\partial  
x^n}}{\frac{\partial}{\partial A_n}}}$ \qquad, \qquad
$X_{x^m} = \displaystyle{{\frac{\partial x^m}{\partial  
A_n}}{\frac{\partial}{\partial x^n}} - {\frac{\partial x^m}{\partial  
x^n}}{\frac{\partial}{\partial A_n}}}$

Furthermore, the inner product of any globally hamiltonian  
vectorfield $X_f$ with the symplectic 2-form of the system $\omega$,  
should result in: $< X_f , \omega > = - df$.
\end{thebibliography}
\end{document}